\documentclass[12pt,preprint]{aastex}

\begin{document}

\title{Is the cosmic UV background fluctuating at redshift $z\simeq 6$ ?}

\author{Jiren Liu\altaffilmark{1,3}, Hongguang Bi\altaffilmark{2},
Long-Long Feng\altaffilmark{1,4} and Li-Zhi Fang\altaffilmark{2}}

\altaffiltext{1}{Purple Mountain Observatory,Nanjing, 210008, P.R. China.}

\altaffiltext{2}{Department of Physics, University of Arizona,
Tucson, AZ 85721}

\altaffiltext{3}{Center for Astrophysics, University of Science
and Technology of China, Hefei, Anhui 230026, P.R. China}

\altaffiltext{4}{National Astronomical Observatories, Chinese Academy of
    Science, Chao-Yang District, Beijing 100012, China}

\begin{abstract}

We study the Gunn-Peterson effect of the photo-ionized intergalactic
medium(IGM)  in the redshift range $5< z <6.4$ using semi-analytic 
simulations based on the lognormal model. 
Assuming a rapidly evolved and spatially uniform ionizing background,
the simulation can produce all the observed abnormal statistical features 
near redshift $ z\simeq 6$. They include: 1) rapidly increase of 
absorption depths; 2) large scatter in the optical depths; 3) long-tailed
distributions of transmitted flux and 4) long dark gaps in spectra. 
These abnormal features are mainly due to rare events, which correspond to the
long-tailed probability distribution of the IGM density field, 
and therefore, they may not imply significantly spatial fluctuations 
in the UV ionizing background at $z\simeq 6$. 

\end{abstract}


\section{Introduction}

Recent observations of the absorption spectra of the highest
redshift quasars show very strong Gunn-Peterson(GP) troughs at $z \simeq
6$ (Fan et al. 2002; Songaila \& Cowie 2002; White et al. 2003;
Songaila 2004; Becker et al. 2006; Fan et al. 2006). The spectra
are found to be dramatically different from low redshift ($2 <
z<4$) samples. First, the blue sides of the Ly$\alpha$ emission
lines are almost completely absorbed, indicating very high
absorption optical depths; and second, there are long flux dark gaps of
about tens of Mpc separated by tiny light leaks in the spectra.
It has been suggested that these observations indicate the end
stage of reionization at redshift $z \simeq 6$.

In low redshift range, $z<4$, Ly$\alpha$ forests can be explained by
absorptions of fluctuated neutral hydrogen atoms in the IGM with a
uniform UV ionizing background (e.g. Rauch 1998). In this model,
about 80\% of the baryons are associated with observed Ly$\alpha$
forest features, while 10\% are in the smaller column densities
which are not detectable and the other 10\% are in collapsed
objects (e.g. Bi \& Davidsen 1997).

If the distribution of the IGM is uniform and the ionizing
background is independent of redshift, the photoionization
equilibrium requires the mean optical depth  $\tau \propto n_{\rm HI}/H(z)
\propto n^2/H(z) \propto (1+z)^{4.5}$, where $n$ is the number
density of the IGM, and $H(z)$ is Hubble parameter. One can then
expect larger Gunn-Peterson troughs at higher redshift. The current
observations show, however, that the mean optical depth actually
underwent a much faster evolution at $z \simeq 6$ than the
$(1+z)^{4.5}$ relation. To explain this fact it was suggested
that the UV background has a strong evolution around $z \simeq 6$.
Moreover, the observed Gunn-Peterson troughs are characterized
remarkably by a large scatter in statistics. For example, some
spectra show complete absorptions while some others show
apparent transmissions. The scatter is proposed to be due to
the existence of a significantly fluctuating UV background
at $z \simeq 6$ (Fan et al. 2006).

Recently, there are a number of simulations or semi-analytical
models aiming at the GP effect at those redshifts (e.g. Razoumov
et al. 2002; Gnedin 2004; Pascho \& Norman 2005;
Wyithe \& Loeb 2005; Kohler et al. 2005; Gallerani et al.
2006). Those studies generally concluded that the UV background or
photoionization rate should be rapidly decreasing toward $z>6$.
However, it is inconclusive on whether a significant inhomogeneity
in the UV background is necessary. Lidz et al. (2006) correctly pointed
out that transmission spectra are sensitive to rare big voids, and
therefore, the observed scatter in optical depths can be produced
from density fluctuations without assuming an inhomogeneous UV
background. However, it is unclear whether the long-tailed distributions
of transmitted flux and long dark gaps can also be explained by their
simulation, because the size of their simulation box is closer or
even smaller than the scales of dark gaps in quasar spectra, which
can be as large as a few tens h$^{-1}$Mpc. Therefore, the problem of
inhomogeneous UV background is still under a cloud.

There are, at least, two other reasons to motivate us to reconsider
this problem. First, the redshift dependence of the mean mass density
is $\rho \propto (1+z)^3$, and then $\delta \rho/\rho \simeq 3 \delta
z/(1+z)$. Thus, a mass density fluctuation $\delta \rho$ on large scales
is approximately equivalent to a variation in redshift or in wavelength
$\delta z$. Therefore once there is strong evolution of $\tau(z)$ on
redshift, i.e., $d\ln \tau(z)/dz \gg 1$, the fluctuations of density
$\delta \rho$ on large scales will yield large variations of $\tau$, even
when the UV background is uniform. Second, when the GP depth is
very large, the only regions which are left with leaking lights
are rare void events in the random field. Those rare events are
described by the long-tailed PDF of the IGM density field. The long tails
in the lognormal model is found to fit well with high order statistics
of the Ly$\alpha$ transmitted flux (Feng \& Fang 2000, Zhan et al.
2001, Feng et al. 2001, 2003, Jamkhedkar et al. 2000, 2003, 2005).
Therefore, it is worth to study the effect of the rare events in the
lognormal model at high redshifts. Because it is a semi-analytical model,
we can take 1-D simulations with spatial ranges as large as a few hundreds
h$^{-1}$Mpc, and resolutions as fine as a tenth of the
Jeans length (a few h$^{-1}$kpc).

\section{Method of simulations}

We simulate Lyman series absorption spectra of QSOs between $z=3.5$
and 6.4 using the same lognormal method as those
for low redshifts $z \simeq 2 - 3$ (e.g., Bi et al. 1995; Bi \&
Davidsen 1997). In
this model, the density field $\rho({\bf x})$ of
the IGM is given by an exponential mapping of the linear density
field $\delta_0 ({\bf x})$ as
\begin{equation}
\rho({\bf x}) = \bar{\rho}_0\exp[\delta_0 ({\bf x}) -  \sigma_0^2/2],
\end{equation}
where $\sigma^2_0=\langle \delta_0 ^2 \rangle $ is the variance of
the linear density field on scale of the Jeans length. In this model,
the peculiar velocity field is also produced by a Gaussian random field,
for which the relevant parameteres are selected to make the
statistical relation between density and velocity fields follow the
predictions of linear fluctuation theory (Bi \& Davidsen 1997; Nusser
\& Haehnelt 1999; Choudhury et al. 2001; Gallerani et al. 2006).

The dynamical bases of the lognormal model have gradually been
settled in recent years. First, although the evolution of cosmic
baryon fluid is governed by the Naiver-Stokes equation, the
dynamics of growth modes of the fluid can be sketched by a
stochastic force driven Burgers' equation (Gurbatov et al. 1989;
Berera \& Fang 1994; Matarrese \& Mohayaee 2002). On the other
hand, the lognormal field is found to be a good approximation of the
solution of the Burgers' equation (Jones 1999). The one-point
distribution of the cosmic density and velocity fields on nonlinear
regime are consistent with lognormal distribution (e.g. Kofman
et al. 1994, Hui et al. 2000, Yang et al. 2001, Pando et al. 2002).
Especially, it has been shown recently that the velocity field of
the baryonic matter of the standard $\Lambda$CDM model is
well described by the so-called She-L\'ev\u{e}que's universal
scaling formula, which is given by a hierarchical
process with log-Poisson probability distribution (He et al. 2006).

As mentioned in \S 1, our purpose is to study whether the observed
Lyman absorptions of QSOs at high redshifts can really rule out
models with a uniform UV background. Therefore, we assume that the
UV background is spatially uniform, contributed by QSOs and active
galaxies, and is represented by an photoionization rate $\Gamma_{\rm HI}$,
which is allowed to change in the redshift range $3.5 < z < 6.4$. The
baryonic gas is heated by the UV background. The thermodynamical
properties of the IGM are actually complicate, because nonlinear
evolution leads to a multi-phased IGM. For a given mass density,
the temperature of the IGM can be different by 1 to 2 orders (He
et al. 2004). Nevertheless, the IGM temperature is related to
mass density by a power law $T \propto \rho^{a}$ in the density range
$\rho/\bar{\rho} <5$, $\bar{\rho}$ being
the mean density (see also Hui \& Gnedin 1997). The neutral fraction
is solved under photoionization equilibrium assumption. To consider 
the peculiar velocity effect and thermal broadening,
we do a convolution of the neutral hydrogen density and velocity 
field with Voigt profile. This yields the absorption optical depth.

We use the standard $\Lambda$CDM cosmological model that is generally
accepted in this type of studies and consistent with the new WMAP data
(Spergel et al. 2006):
$\Omega_{DM} = 0.3$, $h=0.74$, $\sigma_8 = 0.82$, $\Omega_{b} h^2 = 0.025$
and $a=1/3$. The temperature at the mean density is $1.3\times 10^{4}K$
and there is a cut-off of the minimal temperature at $10^4K$.
In redshift space, the size of the simulation samples is given by
$z-0.3$ to $z+0.3$. There are $2^{14}$ pixels in each redshift
range $z\pm 0.3$. We produced $1000$ samples for $3.5<z<5$, and
2000 for $5<z < 6.4$, so as to get the GP depth and their variance.
When comparing to relevant observations, we have divided the samples to
sub-samples with about the same redshift interval (which is 0.15 in
Fan et al. 2006), properly smoothed the mock spectra with observational
resolution and added instrumental noise.

\section{Simulation Results}

1. {\it GP optical depth.} We first try to fit the strong
evolution of the GP optical depth $\tau(z)$ of Ly$\alpha$
absorption. This can easily be done if we take the
photoionization rate to be
\begin{equation}
\Gamma_{\rm HI}(z) = 7\exp\{-[(1+z)/(1+z_0)]^3 \}
\end{equation}
where $\Gamma_{\rm HI}$ is in unit of 10$^{-12}$ s$^{-1}$,
$z_0$ being a free parameter.
Eq.(2) shows that $\Gamma_{\rm HI}(z)$ undergoes a strong evolution when
$z > z_0$, probably due to a strong evolution of star formation rates.
This model is consistent with the general argument that
collapsed objects underwent a fast evolution at larger redshifts (Bi
et al. 2003). We found that $\tau(z)$ in redshift range $3.5<z<6.4$
can be well fitted by eq.(2) if the fitting parameter $z_0$ is in the
range 3.0 - 3.2. The result is shown in Figure 1, and the observed data
are taken from Songaila \& Cowie (2002) and Fan et al. (2006).

The resultant neutral fraction is about $2.2\times 10^{-4}$,
$8.2\times 10^{-4}$ and $2.6\times 10^{-3}$ at $z=5.5, 6.0$ and 6.4
in the $z_0=3.0$ model, and $1.2\times 10^{-4}$, $4.0\times 10^{-4}$
and $1.1\times 10^{-3}$ in the $z_0=3.2$ model.

2. {\it Variance of GP optical depth.} Figure 2 presents the
variance of $\tau(z)$ from an ensemble of 1000 ($3.5<z<5$) or 2000
($5<z<6.4$) simulation samples. They are in good agreement with
observed data (Fan et al. 2006), also shown in Figure 2.
When we look into the details of simulation samples, we found the
scatter is really large. That is to say, the ensemble of simulated
spectra contains samples of complete absorption as well as samples
of apparent transmission. The optical depth is
very sensitive to the density fluctuations when the average GP optical
depth is large. Around $z=6$, the average transmission is basically
contributed by lights leaking from only a few low density voids
in the fluctuating density field. They are rare evens, but they
are not negligible if the PDF of the density is long-tailed.
In other words, the variance of the optical depth mainly depends
on long-tailed low-density events. If a physical phenomenon biases
to rare events, the variance generally is large.

3. {\it PDF of flux.} Figure 3 presents the probability
distributions of the transmitted flux of Ly$\alpha$ absorption at
redshifts $z=5.5$, 5.7 and 6.0. The observed data are from Fan et
al. (2002). Here we use the simulation size of $0.24$ in redshift,
smooth the simulated spectra by a Gaussian
instrument of resolution 2600 and then binned them into pixels of
$35$ km s$^{-1}$, which are about the same as observation. We also
add noises in the simulated spectra. The result shows again that the
simulated distributions of the flux are in good agreement with
observations. Figure 3 shows that most pixels have only very small
transmitted flux, which correspond to opaque regions in spectra.
Figure 3 also shows that although the spectra are general opaque,
there do exist rare but not negligible high transmitted flux.
The long-tail is typical for a lognormal random field.
Figure 3 is consistent with Fig. 2 that the observed large
scatter of the GP optical depth could be caused by the fluctuations of
the IGM density field, not necessarily by irregularities in the
spatial distribution of the UV background.

4. {\it Dark gaps.} In Figure 4, we present the evolution of mean
dark gaps as a function of redshift. The dark gap is defined as
the continuous region in which all pixels have optical depth
larger than 2.5. The mean dark gap from the simulations has a
strong evolution at $z>5.4$, and is generally consistent with observations.
However, we note that the gap at high redshift is sensitive
to the smoothing. If
we use the original spectra without instrumental effects, the
mean dark gap is smaller than that with smoothing. This is because
light leaking areas, which are used to define the boundaries of
the gaps, are very small and show up as
spikes in spectra. Therefore, they will disappear if take a
smoothing on larger scale. Figure 4 shows that the significant
increase of dark gap size at $z\simeq 6$ can also be reproduced
by the strong evolution of $\tau(z)$ alone.

\section{Discussions and Conclusions}

We show that the lognormal model can uniformly and reasonably
explain the observed statistical properties of the Ly$\alpha$ GP
absorption at high redshifts $5 <z < 6.4$, if the intensity of
the uniform UV background, undergoes a strong evolution around $z> 5$.
The large scatter of optical depths and large dark gaps are
resulted naturally from the strong evolution of the optical depth.
These abnormal features may not imply the existence of
significantly spatial fluctuations in the UV ionizing background.
Of course, fluctuations of the UV ionizing background with power
less than the density perturbations are possible in this model.

It is interesting to compare the strong evolution of
the optical depth $(d\ln \tau/dz)\gg 1$, with phase transition.
The condition $(d\ln \tau/dz)\gg 1$ is typical in phase
transition when the IGM transits from the state (phase)
of opaque to transparency with the decrease of the UV background.
It likes phase transition due to the increase of pressure.
Around the point of phase transition, the long wavelength
perturbations will cause large fluctuations in the system considered.
For instance, a normally transparent medium appears milky at critical
point due to the correlations caused by long wavelength
fluctuations. At redshift $z>6$, the density of neutral hydrogen
atoms is high enough, on average, to cause complete Gunn-Peterson
absorption troughs. But long wavelength fluctuations can make some
lines of sight passing through regions where the GP absorption are
low. Although these events are rare, they are not negligible
because the PDF of density distribution is long-tailed.
In this case, the large fluctuations of $\tau$ are directly determined by
$(d\ln \tau/dz)\gg 1$, regardless the inhomogeneity of the UV
background.

We also calculate the statistical properties of Ly$\beta$
and $\gamma$ transmitted fluxes. They have the similar behaviors
as those of Ly$\alpha$. Since the absorption section of Ly$\beta$
and $\gamma$ are smaller, the spectra would be more sensitive to
the UV background. They could be used as better constraints on the
UV background. However, the current data are still not rich enough
to provide an effective comparison with simulation samples.

We conclude that in spite of the Ly$\alpha$ absorption spectra at
$z\simeq 6$ show very different features from those at $z < 4$, the
IGM is probably still in the similar state as $z <4$, i.e., it is
highly ionized by a spatially uniform UV background. The abnormal
statistical features of the Ly$\alpha$ transmitted flux at $z\simeq 6$
are mainly inherited from rare events corresponding to the long-tails
in the PDF of the IGM density field. This picture is
consistent with the other types of observations such as the
luminosity function of Ly$\alpha$ emitter which keeps almost
constant between redshift $z=5.7$ and 6.5 (Malhotra \& Rhoads 2004),
and the latest polarization map of CMB which implies that complete
reionization may have already occurred at $z=7$(Page et al. 2006).
Because non-Gaussianity, such as spiky
structures, intermittency etc. is sensitive to long-tailed events,
our model can be effectively tested with the non-Gaussianity of
the Ly$\alpha$ transmitted flux when more data at high redshifts
become available.

\acknowledgments

We thank our anonymous referee for helpful comments and suggestions.
This work is supported in part by the US NSF under the grant
AST-0507340. LLF acknowledges support from the National Science
Foundation of China (NSFC).

\clearpage



\begin{figure}
\plotone{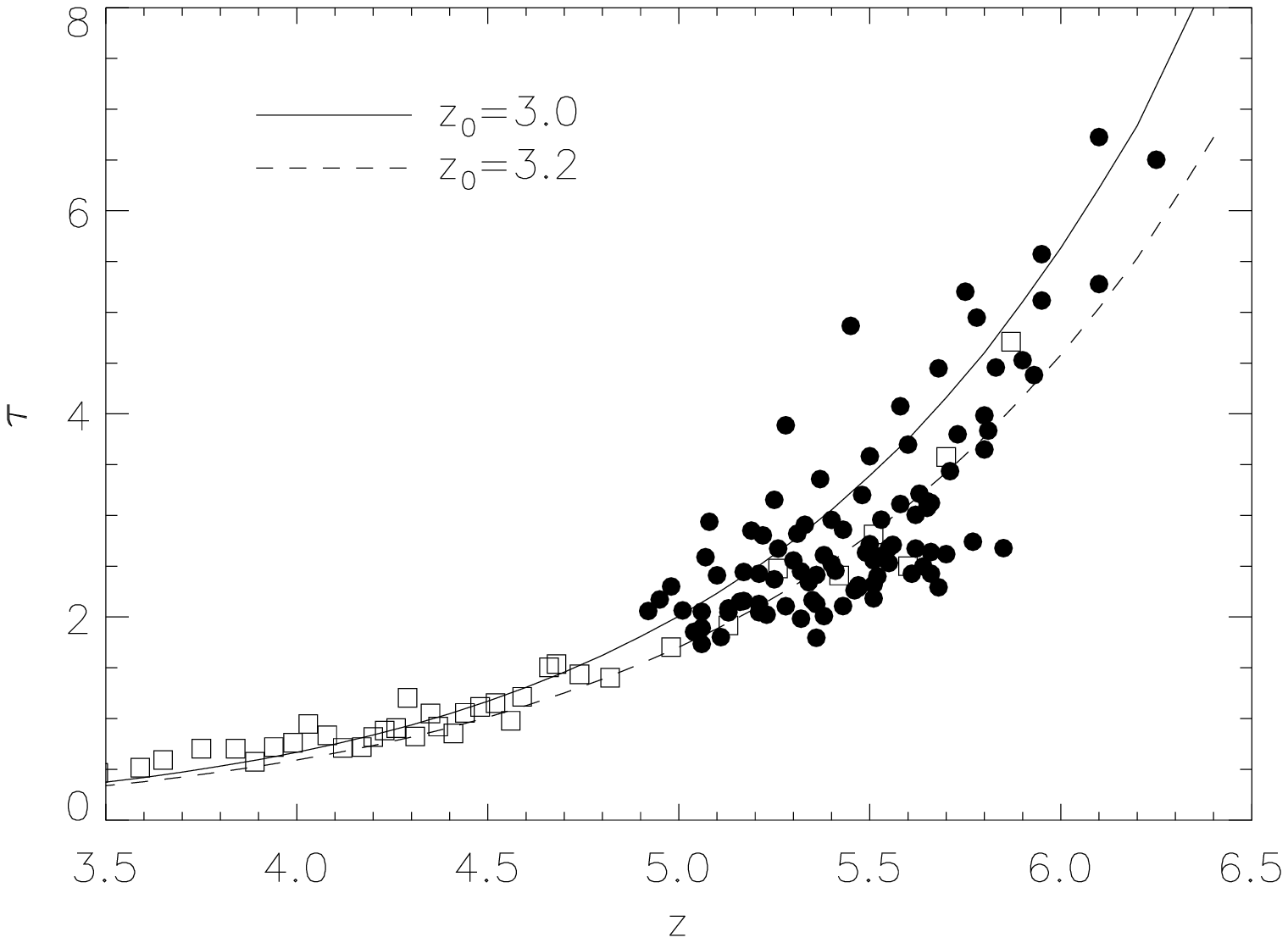}
\caption { $\tau(z)$ vs. $z$ in the redshift
range from
  $z=3.5$ to $z = 6.4$. The observed data  at low redshift are from
Songaila \& Cowie (2002)(square) and $z>5$ from Fan et al. (2006)
(filled dot). The lines are the mean over 1,000 simulation
samples ($z<5$) and 2000 samples ($z>5$). The variance of the
simulation sample is given in Figure 2. The solid line is for
$z_0=3.0$ and the dashed for $z_0 = 3.2$ in Eq. 2.} \label{Fig1}
\end{figure}

\clearpage


\begin{figure}
\plotone{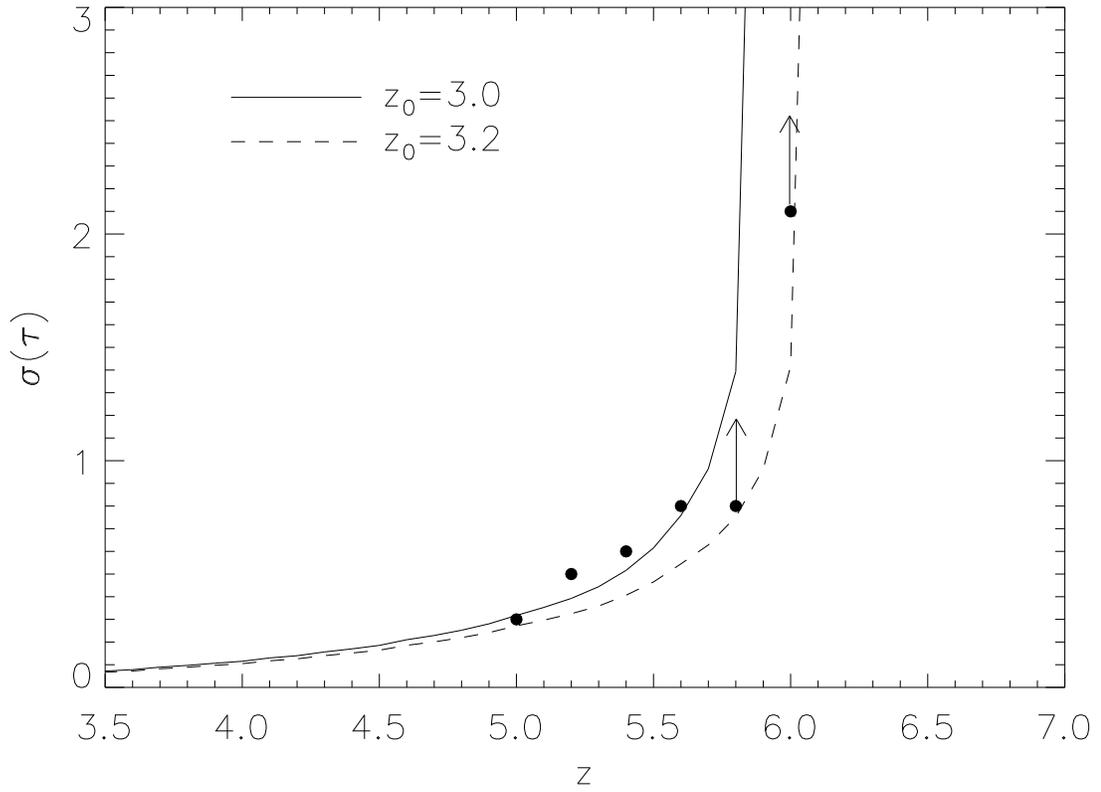}
\caption {The variance $\sigma(\tau)$ as a
function of $z$. The observed data points are taken from Fan et
al. (2006). The simulation samples are the same as Fig. 1.}
\label{Fig2}
\end{figure}

\clearpage


\begin{figure}
\plotone{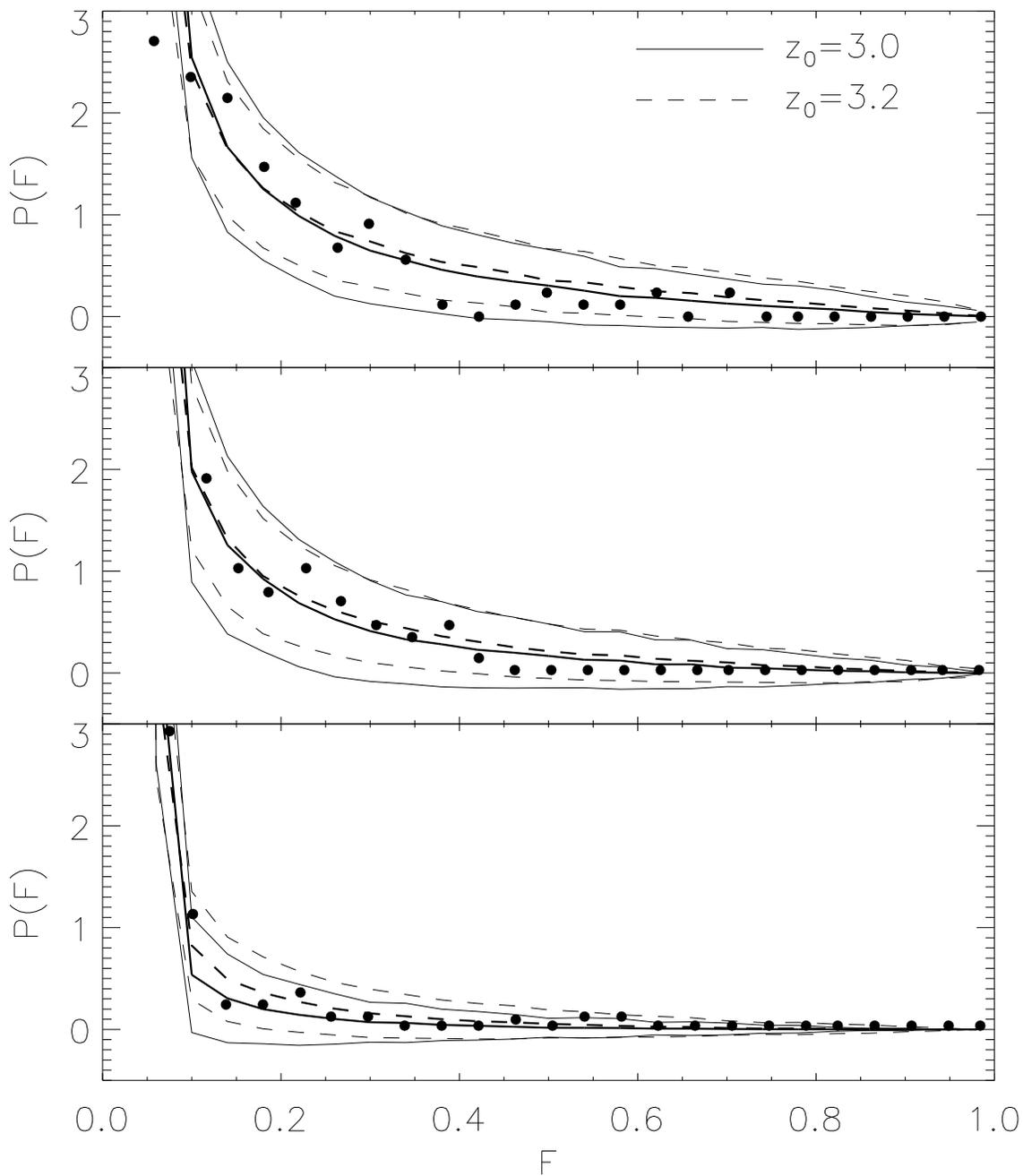}
\caption {Probability distributions of
the transmitted flux at redshifts
 $z=5.5$ (top), 5.7(middle) and 6 (bottom). The observed
data are from Fan et al. (2002). The simulation samples are the
same as Fig. 1. The dark and light solid lines are for $z_0=3.0$
and their 1-$\sigma$ error, respectively. The dark and light
dashed lines are for $z_0=3.2$.} \label{Fig3}
\end{figure}

\clearpage


\begin{figure}
\plotone{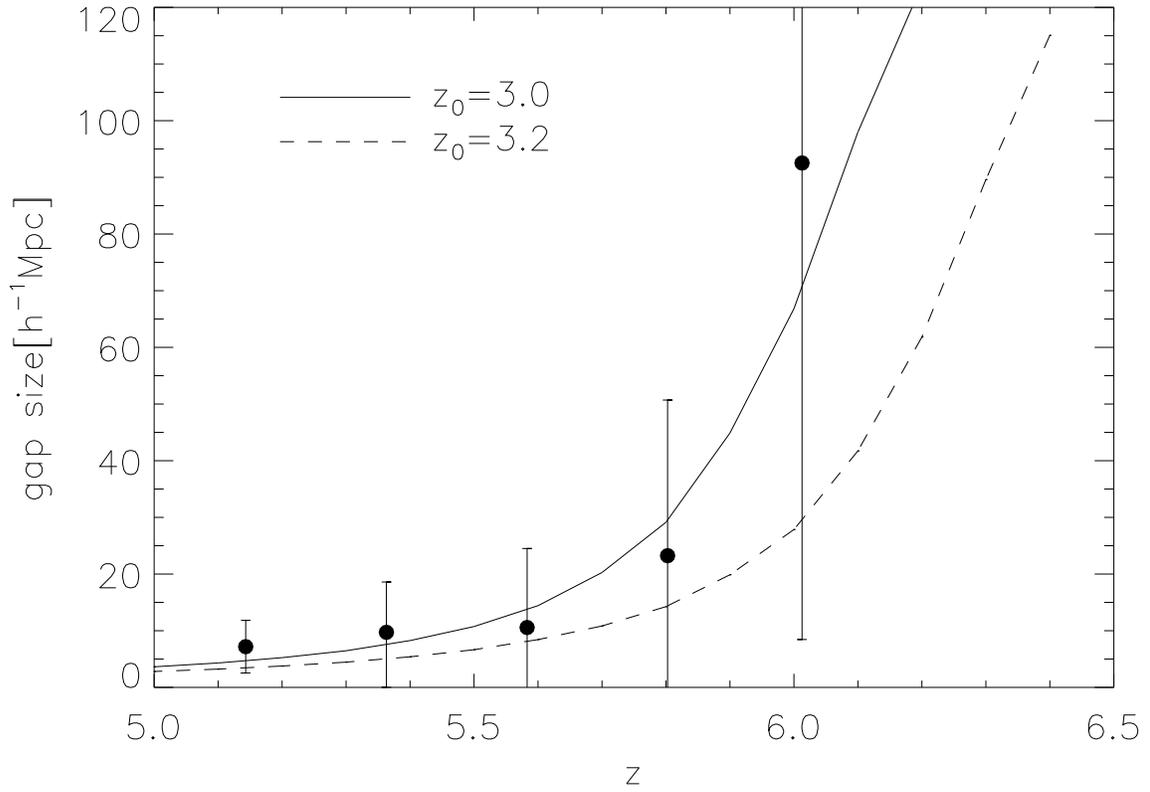}
\caption {The mean size of dark gaps with
$\tau > 2.5$ as function of $z$. The observed data points are from
Fan et al. (2006). The simulation samples are the same as Fig. 1.}
\label{Fig4}
\end{figure}

\end{document}